# The Lyα forest at low redshift: Tracing the dark matter filaments

Patrick Petitjean[1,2], J.P. Mücket[3], and Ronald E. Kates[3]

[1] Institut d'Astrophysique de Paris, 98bis Boulevard Arago, F-75014 Paris, France
[2] UA CNRS 173- DAEC, Observatoire de Paris-Meudon, F-92195 Meudon Principal Cedex, France
[3] Astrophysikalisches Institut Potsdam, An der Sternwarte 16, D-O-1591 Potsdam, Germany

**Abstract.** We study the distribution of low-redshift Lyα clouds in a CDM model using numerical simulations including photoionization and cooling of the baryonic component. The ionizing background is found to be efficient enough to keep most of the gas warm ($T = 1$–$5\ 10^4$ K) and to prevent most of it from collapsing. In this scenario, the numerous low redshift ($z < 0.5$) neutral hydrogen Lyα lines recently revealed by HST observations of QSOs originate in the warm gas that traces the potential wells of filamentary structures defined by the dark matter. The predicted line number density is consistent with observations.

About 25% of the Lyα lines are found to arise in extended envelopes of luminous galaxies or groups of galaxies ($M > 3\ 10^{11}\ M_\odot$) of radius of the order of 1.5 Mpc. The exact number, however, is model dependent. The remainder arises in gas independent of any galaxy and at a distance of 2 to 7 Mpc from the first neighbour galaxy.

## 1. Introduction

A large number of Lyα clouds have been detected very recently at low redshift by HST (Morris et al. 1991, Bahcall et al. 1991, 1993). The observed number density of lines significantly exceeds the simple low-redshift extrapolation of the high-redshift z-dependence. This observation could be explained by a nonzero cosmological constant. However, a significant fraction of the low-redshift population could represent relics of a population of clouds which, at high redshift, were associated with galaxies. Indeed, Lanzetta et al. (1994) have determined the redshift of field galaxies around QSOs observed by HST to search for any association of the Lyα lines with galaxies. They find 11 galaxies coincident in redshift with an absorption system of which nine are Lyα-only systems and two have associated CIV absorptions. They report that (i) most of the $z \sim 0.3$ luminous galaxies are surrounded by gaseous envelopes of radius $\sim 160 h_{100}^{-1}$ kpc and (ii) the fraction of absorption clouds lying at a distance smaller than about $150 h_{100}^{-1}$ kpc from a luminous galaxy is at least 0.35±0.10 and likely 0.65±0.18. However most of the Lyα systems with associated galaxy have $W_r > 0.5$ Å, which translates into $N(\mathrm{HI}) > 5\ 10^{16}$ cm$^{-2}$ for $b = 20$ km s$^{-1}$ (see their Figs. 23 and 26), whereas the bulk of the Lyα lines has $W_r < 0.3$ Å (Morris et al. 1991). Indeed very recent HST observations of bright nearby AGNs by Stocke et al. (1995) have revealed one weak ($W_r = 36$ mÅ) Lyα line with no galaxy within 6 Mpc. The question of whether weak absorptions are associated with galaxies or are completely unrelated to them is still open (Morris et al. 1993).

Galaxies are strongly clustered at low z. If the absorbing gas is associated with galaxies, the absorption lines are expected to cluster on scales smaller than 1000 km s$^{-1}$. The 1D correlation should have an excess on those scales. The issue is unclear however. On the theoretical side, since the size of the Lyα structures may be larger than 10% of the mean separation between absorbers, the spatial correlation function is diluted (Bajtlik 1995). The observational facts are not clearly established as well. At high redshift (z > 2), early investigations have concluded that Lyα lines do not cluster on scales larger than 50 km s$^{-1}$. However, new high resolution data (Cristiani 1995) show that the clustering properties depend on the column density threshold of the sample and a signal is detected in the 1D correlation function of the gas with $N(\mathrm{HI}) > 10^{14}$ cm$^{-2}$ for velocities smaller than 200 km s$^{-1}$. This together with the very recent detection of weak CIV lines in Lyα systems with column densities $N(\mathrm{HI}) \sim 10^{14.5}$ cm$^{-2}$ (Tytler 1995), argues for *part* of the Lyα forest being associated with galaxies. Given the above difficulties, the number of Lyα lines detected at low redshift by HST is not large enough to allow a reliable statistical study of clustering at low redshift. However, detection of clumps of Lyα lines spread over about 1000 km s$^{-1}$ has been reported by Boksenberg (1995).

Since observational evidence for direct association of most Lyα absorptions with bright galaxies is not unequivocal, we suggest here that they arise in gas associated more generally with the potential wells of rich filamentary structures that are ubiquitous in gravitational evolution dominated by dark matter. To obtain a consistent picture within the framework of models for large-scale structure, we study the spatial distribution of Lyα gas and its time evolution using N-body simulations. In this paper we focus our attention on the number density of Lyα lines and the spatial distribution of the Lyα gas at low redshift, in particular on its possible association with very extended disks or halos of galaxies. The properties of these systems at high redshift and their time evolution will be addressed in a forthcoming paper.

## 2. Simulations

We have improved and adapted the particle-mesh (PM) code described in detail by Kates et al. (1991; hereafter KKK) and Klypin & Kates (1991; hereafter KK) to include the effects of photoionization. Let us briefly review the procedure, emphasizing those features relevant to the present investigation:

The PM code simulates the kinematics of the dark matter and the thermodynamics of baryons. Here, we work in a (25 Mpc)$^3$ box (unless otherwise stated, $h = .5$ is assumed) with $128^3$ particles on a $256^3$ periodic grid. The positions, velocities, "temperatures", and gas densities associated with particles are followed starting from a COBE-normalized CDM spectrum and utilizing the Zel'dovich approximation until it breaks down.

The baryonic material is approximated as following trajectories of the dark matter with *constant* $\Omega_b = .1$. This approximation is quite accurate in all regions prior to the first pancakes (or "shocks"). After shocking, the accuracy of the constant-$\Omega_b$ approximation is of course degraded in regions where secondary shocking and/or very efficient cooling take place (e.g., in clusters and in the centres of galaxies, see KK and KKK), but accuracy in these regions is not critical here. The conditions for maintaining constant $\Omega_b$ are well satisfied in Lyα clouds, at least within the mini-halo picture (Rees 1986): The potential wells of mini-halos must be deep enough to prevent the gas from escaping, but still shallow enough to keep the central matter density below levels permitting strong cooling and subsequent star formation to occur.

As perturbations grow, the first objects start to collapse, producing shocks in the gas. At a shock, the temperature assigned to the gas in a particle with velocity **v** entering a region with local velocity **U** is $kT = \mu_H m_H (\mathbf{v} - \mathbf{U})^2/3$, where $\mu_H$ is the molecular weight and $m_H$ the mass of the hydrogen atom. We take primordial abundances throughout: Although as discussed above at least part of the Lyα forest may contain metals, abundances cannot be so high as to significantly influence the quantities of interest here.

After assignment of temperature, we follow the thermal history by integrating the energy equation along particle trajectories, which takes the form

$$\frac{dT}{dt} = (\gamma - 1)\left[\frac{T}{n_H}\frac{dn_H}{dt} - \frac{\mu_M}{\mu_H}\frac{1}{kn_H}(\Lambda_{\rm tot} - \Gamma_{\rm phot})\right] \quad , \quad (1)$$

where $\gamma = 5/3$ and $n_H$ is the number density of hydrogen atoms. Here, $\Lambda_{\rm tot}$ is the total cooling rate (see KK and KKK) including radiative and dielectronic recombination, bremsstrahlung, collisional excitation (Black 1981) and Compton cooling (the latter is the most important process at $z > 5$).

The *heating* rate due to photoionization by the ionizing background (i.e., from QSOs and galaxies) is denoted in Eq.(1) by $\Gamma_{\rm phot}$: The heating rate is computed as the sum of the contributions by HI, HeI and HeII with

$$\Gamma_{HI} = F_o e^{-\tau_{HI}} G \epsilon n_{HI} \quad (2)$$

where

$$G = \int_{\nu_o}^{\infty} \frac{F_\nu}{F_o} \sigma(\nu) d\nu \quad (3)$$

for neutral hydrogen and similar relations for helium. Here, $\sigma$ is the cross-section for photo-ionization, and $\epsilon$ is the mean kinetic energy per photo-ionization (values are taken from Black 1981). The ionizing flux $F_\nu$ is modelled by a power law of index -1. $F_o$ is the ionizing flux at 13.6 eV and depends on redshift. We take $F_o = f 4\pi 10^{-21}$ erg s$^{-1}$ Hz$^{-1}$ cm$^{-2}$ at $z \sim 2.5$, where $f$ is a dimensionless parameter with values varying in the range 0.2-2. Bajtlik et al. (1988) argue that $f$ may be greater than one. On the other hand, from detailed analysis of metal line systems, Petitjean et al. (1992) found that this may well be an upper limit depending on the shape of the ionizing spectrum. A relatively high value $f = 0.9$ is required here in order to reproduce the redshift evolution of the line number density (Mücket et al. 1995).

At redshift $z > 5$, essentially nothing is known about the physical state of the intergalactic medium and the ionizing UV background. In fact, as discussed in KKK and KK, due to the efficiency of Compton cooling at high redshift, the CDM model (and presumably all models with substantial small-scale power) requires some reheating mechanism. Otherwise, most particles shocked prior to $z = 5$ would cool rapidly, implying *excessive* star formation. Here we consider photo-ionization as such a mechanism. The redshift dependence of the ionizing background has been studied by several authors (Bechtold et al. 1987, Miralda-Escudé & Ostriker 1990) and has been shown to be very sensitive to what are the dominant sources of photons. It is clear however that the collapsed regions are the places where the photoionizing photons originate.

To treat this question consistently, we take the ionizing flux intensity proportional to the rate at which material cools below 5000 K in the simulation. This simulates an ionizing flux, produced as a consequence of gas collapsing in the dense regions, and then playing the role of a regulator preventing most of the gas in intermediate density regions from collapsing. The redshift dependence is thus consistently given by the simulation; the only free parameter here is the normalization: We assume $f = 1$ at $z = 5$. The resulting redshift dependence is found to be a power law $(1+z)^\beta$ with $\beta = -2$ and 2 for $z \geq 5$ and $z \leq 2.5$ respectively. Between $z = 2.5$ and 5, the flux is approximately constant. This is very close to what is expected in Models 6 and 8 of Miralda-Escudé and Ostriker (1990).

In Eq. (2), $\tau$ is a characteristic optical depth in the cloud. To estimate it, we model each particle as an independent cloud of constant density taken as the local density given by the code. The radius is thus known from the mass of the particle, and $\tau$ is the optical depth in the centre. This model would tend to overestimate the number of strong lines, since gravity makes the cloud more centrally concentrated; the number obtained here should thus be regarded as an upper limit (see Section 3). The density ratio $n(\mathrm{HI})/n(\mathrm{H})$ is computed self-consistently, taking into account photoionization and collisional ionization. Although *thermal* equilibrium is certainly not reached for $n_\mathrm{H} < 10^{-4}$ cm$^{-3}$ (see eq. 1), *ionization* equilibrium is a good approximation even for low densities (Duncan et al. 1991).

When material cools below 5000 K, small, dense molecular clouds may form. Therefore the corresponding particules are assigned the label "cooled" which is kept for the remainder of the simulation. In order to take into account density variations through the cloud and subsequent evolution, a detailed hydrodynamic description would be required (see e.g. Cen et al. 1994). However specific studies have shown that this is most important for the high-density regions of the clouds and that the low-density regions relevant to our study are only marginally affected (Murray & Petitjean in preparation).

## 3. Results

Fig. 1 shows the projected spatial distribution at $z = 0$ of the dark matter particles in a $25 \times 25$ Mpc$^2$ slice of depth 2 Mpc. The structures seen form the usual network of filaments connected by nodes where galaxies and groups of galaxies are to be found. The neutral hydrogen column density is computed through the slice, and contours of constant neutral hydrogen column density $N(\mathrm{HI}) = 10^{13}$ cm$^{-2}$ are shown. It can be seen that the Ly$\alpha$ gas traces the gravitational potential wells.

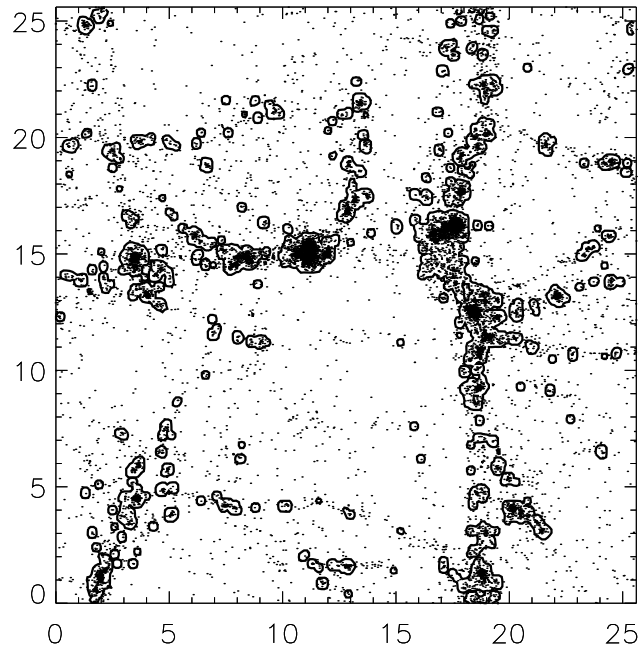

**Fig. 1.** Projected space distribution of cold dark matter particles in a slice of size $25 \times 25 \times 2$ Mpc$^3$ at redshift $z = 0$. Contours for constant neutral hydrogen column density $N(\mathrm{HI}) = 10^{13}$ cm$^{-2}$ are overplotted.

At high redshifts ($2 < z < 3$), the number density $dN/dz$ of Ly$\alpha$ systems is a strongly decreasing function of time, $dN/dz \propto (1+z)^\gamma$. Using high spectral resolution data, Rauch et al. (1992) found $\gamma = 2.1 \pm 0.5$ for $\log N(\mathrm{HI}) > 13.75$ and in the redshift range $2.0 < z < 3.5$. ¿From a compilation of low resolution data, Lu et al. (1991) derived $\gamma = 2.75 \pm 0.29$ for lines with $w_\mathrm{r} > 0.36$ Å. Extrapolating these observations would imply a very low number density at low redshift. Recent observations by HST have shown that the actual observed number of lines is five to ten times larger that what such a simple extrapolation would suggest. The number density of lines per unit redshift is $100 \pm 25$ for $N(\mathrm{HI}) > 10^{13}$ cm$^{-2}$ (Morris et al. 1991) and $15 \pm 2$ for $W > 0.32$ Å (Bahcall et al. 1993) which corresponds roughly (depending also on $b$) to $N(\mathrm{HI}) > 10^{14}$ cm$^{-2}$. In our model the number density of lines with $N(\mathrm{HI}) > 10^{13}$ cm$^{-2}$ is 105 at $z = 0$. The ability to reproduce this number, for a value $f = 1$ within observational constraints on the ionizing flux, represents an important success of the scenario. The number density of lines with $N(\mathrm{HI}) > 10^{14}$ cm$^{-2}$ is about 20 in the model.

The idea that bright galaxies are surrounded by extended envelopes is clearly verified in our simulations. Warm gas is present in large potential wells up to several hundreds of kiloparsecs from the centre (see Fig. 1).

Observational evidence for the association of Ly$\alpha$ gas with halos of galaxies was discussed in the Introduction. In order to quantify the "association" predicted by CDM, we first identify all dark-matter halos in the (25 Mpc)$^3$ simulation box with $M > 3\ 10^{11}\ M_\odot$, obtaining 13 of them. This mass range and spatial density lend credence to the assumption that the halos identified indeed represent individual "bright galaxies" (or groups of galaxies). We then compute the distance $r$ of any cell within which Ly$\alpha$ gas is found with $N(\mathrm{HI}) > 10^{14}$ cm$^{-2}$ to the nearest "bright galaxy" (the number of cells is proportional to the volume occupied by the warm gas). A histogram of $r$ values is shown in Fig. 2. The distribution is essentially bimodal. First, there is a peak at $r < 1.5$ Mpc corresponding to an excess of Ly$\alpha$ gas around the galaxies themselves. This implies, as is apparent in Fig. 1, that bright galaxies are indeed surrounded by halos of radius of the order of 1 Mpc. Secondly there is a broad but well-defined regime out to about 7 Mpc where the distribution drops off sharply which is consistent with a predicted characteristic mean filament separation of about 14 Mpc (Doroshkevitch & Turchaninov, in preparation). In this regime, the Ly$\alpha$ gas is not related to any "bright galaxy". The volume occupied by the gas lying at a distance greater than 1 Mpc to any bright galaxy is twice as large as the volume of the galaxy halos. If we assume, as a first approximation, that the halos around bright galaxies are spherical and the distribution along filaments is cylindrical, the ratio of the projected surfaces of both structures, $S_c/S_s$, is proportional to the ratio of their volumes multiplied by $\frac{8}{3\pi}\frac{R_s}{r_c}$ where $R_s$ is the radius of the sphere and $r_c$ the radius of the cylinder. Taking $R_s \sim 1.5$ Mpc and $r_c \sim 0.8$ Mpc gives $S_c/S_s \sim 3$ and leads to the conclusion that about 25 % of Ly$\alpha$ absorptions should arise within 1.5 Mpc of a bright galaxy. This result is of course sensitive to the operational definition of a "bright galaxy", a long-standing problem in N-body simulations. It is also sensitive to the local description of the Ly$\alpha$ gas. Both questions will be addressed in a forthcoming paper.

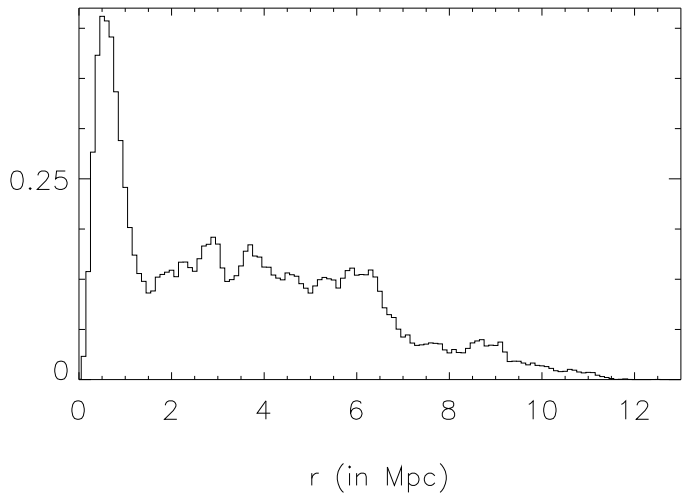

**Fig. 2.** Histogram of the distance from Ly$\alpha$ clouds to the nearest halo of mass $M > 3\ 10^{11}\ M_\odot$

Testing the picture presented here – that the Ly$\alpha$ forest traces the filamentary network of the dark matter – will require new observations. In particular high-quality data should be obtained for the Ly$\alpha$ forest in QSO pairs with very different separations to probe scales from a few tens of kiloparsecs to several megaparsecs. The probability distribution of coincidences at different scales would provide strong constraints on the model. On the other hand, a detailed investigation of the metal content of the Ly$\alpha$ forest at high redshift using 10m class telescopes and at low redshift using shift and stack methods are thus clearly needed.